\documentclass{aa}  

\usepackage{graphicx}
\usepackage{txfonts}
\usepackage[]{xcolor}
\usepackage[]{hyperref}
\begin{document}

   \title{A single-armed spiral in the protoplanetary disk around HD\,34282\,?\thanks{ESO program IDs 096.C-0248(A) and 096.C-0241(A)}}
%An usual spiral in the disk of HD34282
   %\subtitle{I. Overviewing the $\kappa$-mechanism}

\author{J. de Boer\inst{1}
\and C. Ginski\inst{2,1}
\and G. Chauvin\inst{3}
\and F. M\'enard\inst{3}
\and M. Benisty\inst{3}
\and C. Dominik\inst{2}
\and K. Maaskant\inst{4}
\and J.H. Girard\inst{5} 
\and G. van der Plas\inst{3}
\and A. Garufi\inst{6}
\and C. Perrot\inst{7}
\and T. Stolker\inst{8}
\and H. Avenhaus\inst{9}
\and A. Bohn\inst{1}
\and A. Delboulb\'e\inst{3}
\and M. Jaquet\inst{10}
\and T. Buey\inst{7}
\and O. M\"oller-Nilsson\inst{11}
\and J. Pragt\inst{12}
\and T. Fusco\inst{13,10}
}
\authorrunning{de Boer et al.}
\institute{Leiden Observatory, Leiden University, PO Box 9513, 2300 RA Leiden, The Netherlands
\and
Sterrenkundig Instituut Anton Pannekoek, Science Park 904, 1098 XH Amsterdam, The Netherlands\\
\email{c.ginski@uva.nl}
\and 
Univ. Grenoble Alpes, CNRS, IPAG, F-38000 Grenoble, France
\and
~
\and
Space Telescope Science Institute, Baltimore 21218, MD, USA
\and
INAF, Osservatorio Astrofisico di Arcetri, Largo Enrico Fermi 5, I-50125 Firenze, Italy.
\and
LESIA, Observatoire de Paris, Universit\'e PSL, CNRS, Sorbonne
Universit\'e, Univ. Paris Diderot, Sorbonne Paris Cit\'e, 5 place Jules Janssen, 92195 Meudon, France.
\and
Institute for Particle Physics and Astrophysics, ETH Zurich,
Wolfgang-Pauli-Strasse 27, 8093 Zurich, Switzerland
\and
Lakeside Labs, Lakeside Park B04b, A-9020 Klagenfurt, Austria
\and 
Aix Marseille Universit\'e, CNRS, CNES,  LAM, Marseille, France
\and
Max Planck Institute for Astronomy, K\"onigstuhl 17, D-69117 Heidelberg, Germany
\and
NOVA Optical Infrared Instrumentation Group, Oude Hoogeveensedijk 4, 7991 PD Dwingeloo, The Netherlands
\and
DOTA, ONERA, Université Paris Saclay, F-91123, Palaiseau France
}

   \date{Received September 15, 1996; accepted March 16, 1997}

% \abstract{}{}{}{}{} 
% 5 {} token are mandatory
 
  \abstract
{During the evolution of protoplanetary disks into planetary systems we expect to detect signatures that trace mechanisms such as planet-disk interaction. Indeed, protoplanetary disks display a large variety of structures in recently published high-spatial resolution images. However, the three-dimensional morphology of these disks is often difficult to infer from the two-dimensional projected images we observe.}
{We aim to detect signatures of planet-disk interaction by studying the scattering surface of the protoplanetary disk around
HD 34282.}
{We spatially resolve the disk using the high-contrast imager VLT/SPHERE in polarimetric imaging mode. We retrieve a
profile for the height of the scattering surface to create a height-corrected deprojection, which simulates a face-on orientation.}
    %Results
{The detected disk displays a complex scattering surface. An inner clearing or cavity extending up to r $<$ 0\farcs28 (88 au) is surrounded by a bright inclined (i = 56\degr) ring with a position angle of 119\degr. The center of this ring is offset from the star along the minor axis with 0\farcs07, which can be explained with a disk-height of 26 au above the mid-plane. Outside this ring, beyond its
south-eastern ansa we detect an azimuthal asymmetry or blob at r $\sim$ 0\farcs4. At larger separation, we detect an outer disk structure that can be fitted with an ellipse, compatible with a circular ring seen at r = 0\farcs62 (= 190 au) and height of 77 au. After applying a
height-corrected deprojection we see a circular ring centered on the star at 88 au, while what seemed to be a separate blob and outer ring, now both could be part of a single-armed spiral.}
  % conclusions
{We present the first scattered-light image of the disk around HD 34282 and resolve a disk with an inner cavity up to
r $\approx$ 90 au and a highly structured scattering surface of an inclined disk at large height: $H_{\rm scat} / r =$ 0\farcs29 above the midplane at the inner edge of the outer disk. Based on the current data it is not possible to conclude decisively whether $H_{\rm scat} / r$ remains constant or whether the surface is flared with at most $H_{\rm scat} \propto r^{1.35}$ , although we favor the constant ratio based on our deprojections. The height-
corrected deprojection allows a more detailed interpretation of the observed structures, after which we discern the first detection of a
single-armed spiral in a protoplanetary disk.}


   \maketitle
%-------------------------------------------------------------------
%Introduction
%-------------------------------------------------------------------

\section{Introduction}

The study of planet formation in protoplanetary disks has
reached a new era with the possibility to image these disks at
high angular resolution. The developments at both the optical
and sub-mm regime have yielded improvements both in obtainable resolution and in sensitivity. In the visible and Near InfraRed (NIR), the extreme Adaptive Optics (AO) high-contrast
imagers Gemini Planet Imager \citep[Gemini/GPI, ][]{Macintosh2014} and the Spectro-Polarimetric High-contrast Exoplanet
REsearch \citep[VLT/SPHERE,][]{Beuzit2019} instrument reach
close to diffraction limited resolutions of $\sim$50\,milliarcsecond
(mas). These visible and NIR imagers detect light that is scattered by $\sim$micron ($\mu$m) sized dust grains in the disk surface.
Long baseline observations at sub-mm wavelengths with the Atacama Large Millimeter Array (ALMA) trace thermal emission
of gas and $\sim$mm sized dust grains in disks at similar ($<$\,0.1\arcsec{})
resolutions \citep{HLTauALMA2015}.\\
Previous assumptions of smooth continuous disks are challenged by high-resolution images with the detection of spiral
arms \citep{Muto2012,Grady2013,Benisty2017},
gaps and rings \citep{Quanz2013,HLTauALMA2015,deBoer2016}. While the presence of planets is often considered to be the cause for such structured disks \citep{Ogilvie2002,Dong2016,deJuanOvelar2013}, multiple explanations exist for the different types of structure.
Among other explanations, disk gaps and cavities can also be
caused by the presence of dead zones \citep{Flock2015,Pinilla2016} or photoevaporation \citep{Alexander2006}, while
spiral arms can be due to self-gravity in the disk \citep{Lodato2004,Dipierro2015} or due to temperature gradients caused by shadowing \citep{Montesinos2016}.\\
To date, almost exclusively double- (or multiple-)armed spiral structures
have been detected in protoplanetary disks.\footnote{An exception may be the the disk around V1247\,Ori observed by \cite{Ohta2016}, who do mention the possibility of a single spiral arm based on their data. }
Double-armed spirals typically display very similar contrasts between both arms and the surrounding disks\footnote{With the notable exception of stellar fly-by induced structures, see e.g. \citealt{Cuello2020, Menard2020}.}. Self-gravity and shadowing can readily explain such double-armed spirals. However, spirals induced
by a (proto)planet are expected to be either single-armed (outside of the planet location) or double-armed (\citealt{Zhu2015,Dong2015a,Miranda2019}). Currently, no protoplanetary disk has been detected with a confirmed single-armed spiral structure. When new pre-main sequence systems are imaged at high resolution, nearly all appear to have distinct particularities. Encountering new disk features allows us to study the various details of the interactions in evolutionary processes, which will ultimately
yield a better understanding of the general principles driving disk
evolution and the formation of planetary systems.\\
HD\,34282 (alias V1366\,Ori) is an interesting candidate to
search for structure in the protoplanetary disk surrounding this
Herbig Ae star \citep[][Spectral type: A3V,]{Merin2004}. 
\citet{Merin2004} determine the stellar age at 6.4$\pm$0.5\,Myr and mass 
$M_\star$=1.6$\pm$0.3\,$M_\odot$, assuming a distance of 348$^{+129}_{-77}$\,pc. The second data
release (DR2) of GAIA \citep{GAIA-DR2-2018} constrains the
distance to 312$\pm$5\,pc. Based on Gaia DR2, \citet{Vioque2018}
adjust the stellar age to 6.5$^{+2.4}_{-0.6}$\,Myr and $M_\star$ = 1.45$\pm$0.07\,$M_\odot$.\\
The presence of a Keplerian disk surrounding HD\,34282 is
inferred from its strong IR excess \citep{Sylvester1996} and
double peaked CO\,(J = 3-2) emission \citep{Greaves2000}.
\citet{Pietu2003} resolved an inclined ($i = 56\pm3^\circ$) disk with the
IRAM interferometer, and determined a temperature law compatible with a flaring disk heated by the central star. \citet{Khalafinejad2016} have used the Spectral Energy Distribution (SED)
and marginally resolved Q-band images to create a radiative
transfer model of the disk, which predicts a gap of 92$^{+31}_{-17}$\,au (at 348\,pc, which converts to 82$^{+28}_{-15}$\,au at 112\,pc). Indeed, a 0.24\arcsec{}
(= 75 au) wide cavity is detected in the ALMA band 7 continuum image of \citet{vanderPlas2017}. In the continuum image with a spatial resolution of 0.1\arcsec{}$\times$0.17\arcsec{}, the disk is resolved and detected up to 1.15\arcsec{} from the star as a $\sim$60$^\circ$ inclined ring. Furthermore, the ring contains a "vortex-shaped" asymmetry near
the south-eastern ansa of the elliptical ring.\\
\citet{Maaskant2014} list the most famous Herbig Ae/Be
stars in order of increasing flux ratio $F_{30\,\mu\mathrm{m}} /F_{13.5\,\mu\mathrm{m}}$, which can
be used as a tracer for disk flaring. HD\,34282 is placed among
the top three most flared disks, below IRS\,48 and HD\,135344B.
Based on the predictions of a large gap and very strong flaring
of the disk surface, we expect that HD\,34282 harbors an easily
detectable disk with clearly resolvable features.\\
In this study, we have used VLT/SPHERE to spatially resolve
the scattering surface of the protoplanetary disk of HD\,34282.
Our SPHERE observations are listed in Sec. 2, followed by a
description of the reduction and post-processing of this data in
Sec. 3. We present our results and perform a geometrical analysis
of the images to reconstruct the disk morphology in Sec. 4. We
finish with a brief discussion in Sec. 5 and the conclusions in
Sec.6.
%--------------------------------------------------------------------
%Observations
%--------------------------------------------------------------------

\section{Observations}

We observed HD 34282 with VLT/SPHERE’s InfraRed Dual-
beam Imager and Spectrograph (SPHERE/IRDIS, \citealt{Dohlen2008}) and Integral Field Spectrograph (SPHERE/IFS, \citealt{Claudi2008}) instruments, using the apodized pupil Lyot coronagraph (ALC\_YJH\_S, \citealt{Boccaletti2008, Martinez2009})
with a diameter of 185 mas and inner working angle of 100 mas
(Wilby et al. in prep.).\\
As part of the Guaranteed Time Observations (GTO) of the
Disk group within the SPHERE consortium, we have observed
HD 34282 with IRDIS, using the Dual-beam Polarimatric Imaging mode (IRDIS/DPI, \citealt{Langlois2014, deBoer2020, vanHolstein2020}) in broad-band J filter ($\lambda_0$ = 1.258\,$\mu$m, $\Delta\lambda$ = 197\,nm). The IRDIS/DPI observations are recorded on December 19 of 2015 with Detector Integration Times (DITs) of 64 s in field-stabilized mode. During the total exposure time of 94 min, we cycled through the four
half-wave plate positions (0$^\circ$ , 45$^\circ$, 22.5$^\circ$ and 67.5$^\circ$, to modulate the linear polarization components) eleven times.\\
During the GTO of SHINE (SpHere INfrared survey for Exoplanets) on October 25 of 2015, we have observed HD 34282 in
IRDIFS mode: The IFS recorded in Y-J band ($\lambda$ = 0.96 - 1.34\,$\mu$m,
with spectral resolution $\Delta\lambda$ = 55.1\,nm); IRDIS in Dual-band
Imaging mode (DBI, \citealt{Vigan2010}) with the H2-H3 filter
combination (H2: $\lambda_0$ = 1.5888\,$\mu$m, $\Delta\lambda$ = 53.1\,nm, H3: $\lambda_0$ = 1.6671\,$\mu$m, $\Delta\lambda$ = 55.6\,nm). The SHINE observations with both IRDIS and IFS were recorded in pupil-stabilized mode, using DITs of 64\,s for a total exposure time of 68\,min, during 54$^\circ$ of field rotation.

\begin{figure*}
\center
\includegraphics[width=0.98\textwidth]{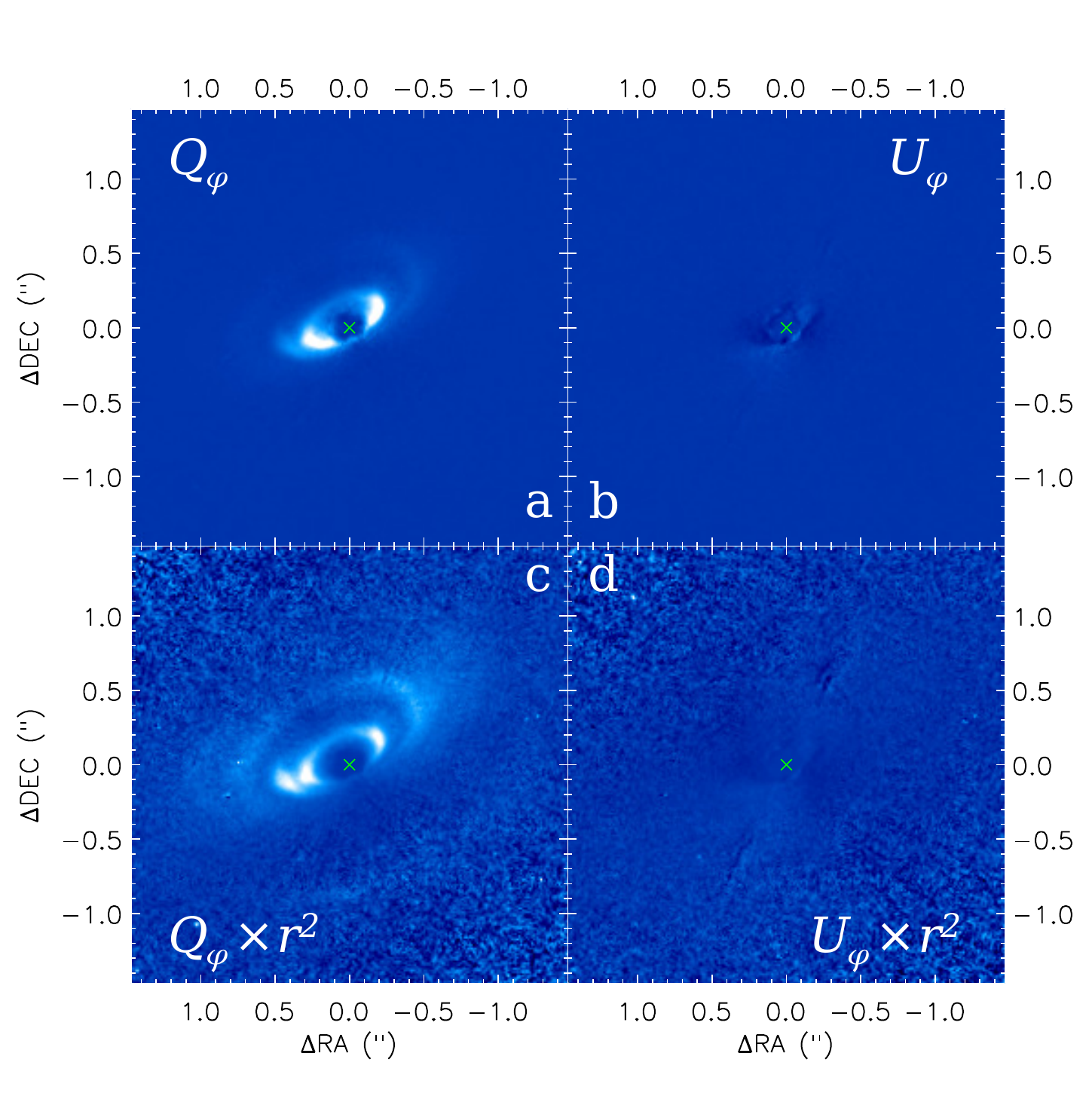} 
\caption{VLT/SPHERE/IRDIS polarimetric images of HD 34282. In all panels, north is up, east is left, and the star-center is annotated with a green
cross. All images are shown on a linear color map. [a:] The Q$_\phi$ image shows inner cavity, surrounded by the bright inner edge (or wall) of the outer disk as an ellipse with its center slightly
offset from the star in north-eastern direction. Faint structures are detected directly outside this inner wall. [b:] The U$_\phi$ image shown with the
same intensity range as panel a. [c:] To highlight the outer disk structures we scaled the Q$_\phi$ image with the disk radius squared, corrected for the
inclination \citep[][56\degr]{Pietu2003}. To a first order, this r$^2$ scaling accounts for the decrease in stellar irradiation of the disk. Because we did not
take the height of the scattering surface into account we show this scaling for illustrative purpose only: to highlights the fainter structures outside
the inner wall. [d:] U$_\phi$ with the same r$^2$ scaling and displayed for the same intensity range as panel c.
} 
\label{fig:hd34282-sphere-reduced}

\end{figure*}

%--------------------------------------------------------------------
%Data reduction
%--------------------------------------------------------------------

\section{Data reduction}

We reduced the data using the method discussed in detail by \cite{deBoer2016}. The linear polarization images are retrieved
by computing Stokes vector components Q and U with the double difference method. Next, we determine for each pixel the angle $\phi$ between the meridian and the line crossing both this pixel
and the star-center (increasing from north to east). We use the
Stokes vector components to determine where the polarization
angle at each point in the image is aligned in azimuthal direction
with respect to the star-center (Q$_\phi$ , negative signal represents radial alignment of the polarization direction) and where the polarization angle is aligned $\pm$45$^\circ$ with respect to the star-center (U$_\phi$) according to (\citealt{Schmid2006}):
\begin{eqnarray}
Q_\phi = -Q\,\mathrm{cos}(2\phi) - U\,\sin (2\phi),\\
U_\phi = -Q\,\sin (2\phi) + U\,\mathrm{cos}(2\phi). 
\end{eqnarray}

In disks seen at near face-on orientation (e.g., TW Hydrae,
\citealt{vanBoekel2017}) we can expect the U$_\phi$ signal to be dominated by noise and instrumental or reduction artefacts. However, a study based on radiative transfer modeling by \cite{Canovas2015} showed that due to breaking of symmetry (of scattering angles) in protoplanetary disks seen at high inclination angles, multiple scattering can cause a true deviation from azimuthal polarization: i.e., U$_\phi$ disk signal. Although the inclination of the disk around HD 34282 is known to be high (56\degr - 60\degr), we used the assumption that the disk signal is dominated by single scattering to correct for the reduction artifacts caused by an
insufficient instrumental polarization correction. We removed a
constant scalar multiplied with the total intensity image from the Q and U images independently that yielded the lowest absolute signal in the U$_\phi$ image along an annulus centered on the star. On our final images, we corrected for bad pixels by applying a sigma filter. For each pixel in our frame, we measure the standard deviation in a box surrounding (but excluding) this pixel with width of seven pixels. When the central pixel deviates by more than 3$\sigma$ from the mean of the remaining pixels within the box, it is replaced by this mean value.\\ In accordance with \cite{Maire2016}, the resulting Q$_\phi$ and U$_\phi$ images are aligned with true north by rotating them with 1.8$^\circ$ in clockwise direction and a pixel scale of 12.26\,mas is assumed.

%--------------------------------------------------------------------
%Results and geometrical analysis
%--------------------------------------------------------------------

\section{Results and geometrical analysis}

\begin{table}
	\centering
	\caption{Ellipse parameters for the fits to the features R1 and R2 as
shown by the green ellipses in Fig. 2. Since our fitting routine does not
determine the systematic errors, we round all values conservatively in
the third column and do the same for the 1$\sigma$ random errors displayed in
column 4.}
	\label{tab:ellipse-fit}
	\small
	\begin{tabular}{clll}
	Ring & Parameter & PDI-$J$ & $\sigma$ \\
	\hline
	\hline
	R1 & Semi-major axis (\arcsec{}) & 0.62 & 0.01 \\
	   & Semi-minor axis (\arcsec{}) & 0.34 & 0.01 \\
	   & RA offset $u_x$ (\arcsec{}) & 0.10 & 0.01 \\
	   & Dec offset $u_y$ (\arcsec{}) & 0.17 & 0.01 \\
	   & Offset angle ($^\circ$) & 31 & 1 \\
	   & Position angle ($^\circ$) & 118 & 1 \\
	   & Inclination angle ($^\circ$) & 57 & 1 \\	
	   & $H_{\rm scat}$ (au) & 74 & 1 \\
	   & $H_{\rm scat}/r$ (au) & 0.37 & 0.01 \\
	\hline
	R2 & Semi-major axis (\arcsec{}) & 0.28 & 0.01 \\
	   & Semi-minor axis (\arcsec{}) & 0.16 & 0.01 \\
	   & RA offset $u_x$ (\arcsec{}) & 0.04 & 0.01 \\
	   & Dec offset $u_y$ (\arcsec{}) & 0.06 & 0.01 \\
	   & Offset angle ($^\circ$) & 36 & 1 \\
	   & Position angle ($^\circ$) & 119 & 1 \\
	   & Inclination angle ($^\circ$) & 56 & 1 \\	
	   & $H_{\rm scat}$ (au) & 26 & 1 \\
	   & $H_{\rm scat}/r$ (au) & 0.28 & 0.01 \\
	\hline
		
	\end{tabular}
\end{table}

We show the final Q$_\phi$ and U$_\phi$ images in Figs. 1a and 1b. To highlight the outer disk structures, we have multiplied Q$_\phi$ and U$_\phi$ in
Figs. 1c and 1d with the square of the separation (r$^2$) to the star, using inclination (i = 56\degr) and the assumption that the scattering surface has no height above the mid-plane (thin disk). We realize that the thin-disk assumption is an unlikely simplification. Hence, the inclination corrected r$^2$ scaling is only used for visualization purposes: to simultaneously show the inner and outer disk features. A similar effect of covering a larger dynamical
range in an image is reached by showing them in logarithmic
scale, as we do in Fig. 2. The log scale has the benefit that is unbiased by our choice for the inclination and height of the scattering surface, but is unsuitable for a clear display of the U$_\phi$ image (which can have large negative values) and shows the outer regions more diffusely than the r$^2$-scaled image. \\
Fig. 2 is annotated with the most distinct disk features. At
the smallest separations, we detect a clearing in the disk or inner cavity, which is surrounded by a ring feature (R2). Further
from the star we detect a second ring-like feature (R1), an arc
(A1) below a dark lane in the southwest, an arc (A2) outside the
southeastern ansa of R1 and diffuse signal on the opposite side
(A3, outside the northwestern ansa of R1), and finally a bright
arc or blob (B1) in-between the southeastern ansae of R1 and R2.\\
Previous SED analysis by \cite{Maaskant2014} already
shows that the disk is not geometrically thin and most likely flared. Additionally, the disk images of Figs. 1a and 1c show indications that the thin
disk assumption is unlikely to be correct. Therefore, all quantitative analyses below are performed on the non-scaled image
(Fig. 1a). The first indication in our data that the thin-disk assumption is incorrect can be derived from the faint arc A1 and
the dark lane in between A1 and R1, which strongly resemble
the backward facing (or bottom) surface and the obscured mid-plane of an optically thick disk, respectively. The second hint that this disk is thick can be found by comparing the centers of the innermost ring R2 (the inner scattering wall) and the outermost ‘ring-like’ feature R1 to the position of the star (green X in Figs. 1 and 2). The rings are clearly offset towards northeastern direction (along the disk minor axis) with respect to the star-center. When we use the assumption that the rings are in fact circular and centered around the star (when observed face-on),
\cite{Lagage2006} and \cite{deBoer2016} have shown that for ring R, we can explain
the apparent offset $u_\mathrm{R}$ of that ring detected at an inclination ($i$) as a projection caused by the ring’s height of the scattering surface
$H_{\rm scat,R}$ above the mid-plane:
\begin{eqnarray}
u_{\rm R} = H_{\rm scat,R}\,\sin\,i.
\end{eqnarray}
Therefore, we can directly determine $H_{\rm scat}$ for R1 and R2 by measuring the offsets of the ring features.\\
Note that the previous interpretation for the ring offsets implies that the south-west is the near (or forward scattering) side of the disk, which is in agreement with feature A1 being the bottom side of the disk and the dark region between R1 and A1 being the obscured mid-plane of the disk. All features are brightest roughly along the major axis of the disk. Along this axis, the scattering angle is close to 90\degr , which is typically where we expect scattering to cause the highest degrees of linear polarization.

\subsection{Ellipse fitting}

To measure the offset of the ring-centers with respect to the star, we fit ellipses to R1 and R2 with the method discussed in \cite{deBoer2016}. The ellipse-fit parameters are listed in Table 1. Note that the position angle (PA) along which the ellipses
are offset with respect to the star (listed as ‘Offset angle’, which
increases from north to east) are within 7\degr{} from perpendicular to
the PA of the ellipses, which is a prerequisite for the offset to be
caused by the projection of inclined circular rings that reside on
a plane $H_{\rm scat}$ above the mid-plane. The offset of the inner ring
R2 shifts its south-western (near) side within the coronagraph’s
inner working angle, which makes the detection of signal at this
part of the ring tentative. However, all other regions of this ring
(e.g., the far side and along the major axis) are well outside this
inner working angle. Therefore we have confidently detected the inner edge or wall of the outer disk. We find that the inner wall
lies at $\sim$88\,au, which is in remarkable agreement both with the
inner disk radius of 82$^{+28}_{-15}$\,au, as predicted by \cite{Khalafinejad2016} based on radiative transfer modeling of marginally resolved Q-band data, and with the inner edge of the sub-mm annulus detected at 75\,au with ALMA by \cite{vanderPlas2017}.
To determine the radial profile of the disk height $H_{\rm scat}$ (only
possible to the first order, because we consider the profile to be
smooth and continuous and ignore the possible presence of gaps
in the disk), we extrapolate the height at both rings to all radii by
fitting a power law to the two data points:
\begin{eqnarray}
H_{\rm scat}(r) = 0.06\,r^{1.35\pm0.08},
\end{eqnarray}
with $H_{\rm scat}$ and $r$ in au.
This power law falls between the values that were recently determined for two ringed systems: the non-flaring disk ($\propto$r$^1$) RX\,J1615.3-3255 (\citealt{deBoer2016}) and the strongly flaring ($\propto$r$^{1.73}$) HD 97048 (\citealt{Ginski2016}), and just above the average value ($\propto$r$^{1.2}$) for five disks determined by \cite{Avenhaus2018}.\\
It should be noted that the errors mentioned in Tab.1 are
merely the random errors for a fitting routine that only accepts an ellipse. However, a closer look at Fig. 2 shows that the outer ring R1 does not trace the fitted ellipse as well as R2. This deviation from a ring structure is best visible when one notices the asymmetry between the two ansae of the ellipse: locally more disk signal comes from regions just inside the green ellipse in the northwestern ansa, while on the southeastern ansa most signal is detected outside the fitted ellipse. When R1 is in fact not a ring, the previously determined ellipse and flaring parameters are no longer valid. Therefore, we consider two scenarios as the extreme possibilities: 1) R1 is a ring (albeit irregularly shaped
and/or illuminated) and the surface profile is strongly flared according to Eq. 4. 2) R1 is not a ring, and we choose the flattest geometry that still allows R1 to be irradiated by the central star (i.e., no flaring), which is constrained by the height of R2:
\begin{eqnarray}
H_{\rm scat}(r) = 0.29\,r.
\end{eqnarray}

\subsection{Height-corrected deprojection}
If we deproject the disk image to a face-on orientation, we can
corroborate (or invalidate) that the disk is flaring according to Eq. 4,
because this scenario should yield two concentric circles after
deprojection. Height profiles as given by Eqs. 4 \& 5 allow us to
determine $H_{\rm scat}(r)$ and subsequently determine for each pixel in
the image a corresponding ellipse with offset $u(r)$, using Eq. 3. We use these height maps to create a height-corrected deprojection, as illustrated in Fig. 3a for the non-flaring scenario. First, each pixel is shifted with $-u(r)$, as found for the ellipse containing this pixel (Figs. 3b for the height-map and 3e for the disk
image). This process moves the surface features to the disk mid-plane (effectively creating a ’thin disk’ or flattened image), which yields the most appropriate
image to compare with ALMA images of the mid-plane of the
disk (\citealt{vanderPlas2017}). For simplicity, we have not accounted for signal originating from the bottom of the disk, such as feature A1, in which case these regions are shifted in the same direction ($+u$) as their true offset. This moves signal from the bottom side of the disk even further away from the center, which is convenient, as it avoids confusion between signal originating from the top and the bottom. In the near or forward scattering side of the disk (south-west of the star, but very close to the inner cavity), we are strongly under-sampled: each new pixel on this side of the disk does not individually have a unique corresponding pixel in the original image. This point becomes clear when comparing the contours of the height maps in Figs. 3a and 3b:
the contours in the southwestern side of the disk lie further apart in the flattened map (b) than in the height map of the original image (a). To account for this under-sampling, we resampled the original image with $3\times3$ subpixels, and drew the pixel values for the thin disk image from pixels shifted by $3u$ in the original but resampled image. Figs. 3c and 3f show the final step where we
‘stretch’ or resample the flattened image along the minor axis by increasing the number of pixels in this direction with the factor $1/ \mathrm{cos}(i)$, which produces the height-corrected deprojections for the height-map and the disk image, respectively.\\ 
We investigated this height-corrected deprojection for two $H_{\rm scat}$ profiles: the non-flaring surface of Eq. 5 and the flaring surface of Eq. 4 in Fig. 4, panel a and b, respectively. The south-west (PA $\sim230^\circ$) of panel b is plagued by the fact that the value for multiple pixels are drawn from the same point in the original image, because the height profile of the near side of the inclined power law creates a stronger under-sampling than for the non-flared profile. The green dashed line shows a circle with a radius equal to the semi-major axis of R2 from Tab. 1, which traces the local maxima in the disk surface brightness just outside the cavity in both panels. This inner rim does not show a ring with
a homogeneous brightness distribution for each azimuth angle.
In appendix A we show that we do expect such asymmetries for
this ring even when it is illuminated homogeneously, due to variations in the scattering phase-function with azimuth angle.
The orange dashed line in Fig. 4 a \& b shows a circle with
the radius based on the semi-major axis of R1 from Tab. 1. As expected for the non-flared height correction of Fig 4 a, where
R1 is not assumed to be a ring, the orange circle does not trace
the outer disk feature very well. Roughly between $-45^\circ \leq PA
\leq 45^\circ$ , the orange circle coincides with the local maxima in the
disk, while for $45^\circ \leq PA
\leq 135^\circ$ and $225^\circ \leq PA
\leq 315^\circ$ the
local maxima lie outside and inside of the orange circle, respectively. For the remaining PAs in the south it is rather difficult to
associate and compare any local maxima with the circle-shape.
In panel b of Fig. 4, where we do make the assumption that R1
is a ring, the local maxima do indeed lie closer to the orange circle. However, we can discern a similar trend of the local maxima
lying inside the orange circle in the western side and outside of
this circle on the east, albeit much more subtle than for panel a.
Based on the deprojections for the two extreme height-profiles
we conclude that the outer disk is most likely not shaped like a
ring.

\subsection{Single-armed spiral}
\label{sec: spiral-description}

The deviation of the outer disk feature from a circular shape as
we show in Fig. 4 shows a general trend (i.e. moving outwards in
counter-clockwise direction) that is consistent between the two
extreme possibilities for the height profile. Although we cannot
fully dismiss that the outer region is circular in panel b, it is more
likely that the structure does in fact behave according to a spiral
pattern.\\
In Fig. 5b, we overlay a simple (single-armed) Archimedean
spiral (outer green-white-orange dashed line) on the deprojection
using the non-flaring height profile. The spiral can be described
by
\begin{eqnarray}
r = c_0 + c_1 \phi,
\end{eqnarray}
where $\phi$ is the azimuth angle in degrees, $c_0 \sim 180$\,au and $c_1
\sim 0.2$\,au degree$^{-1}$ . For $r \gtrsim 200$\,au this spiral traces the local maxima better than the orange dashed circle in Figs. 4a and 4b. The
green dashed line traces the brightest regions and winds outward
in counter-clockwise direction from a PA of $\sim 270^\circ$ for another
$270^\circ$. White dashed lines show parts where the signal-to-noise
ratio is too low to determine the local maxima, but are shown
to illustrate that the outer (green-white-orange) and inner (red-
white-red) arms can be described with only two connected, yet
distinct spiral patterns. \\
The orange dashed line (PA$\sim90^\circ$) reconnects the spiral to the local maxima of feature A2. The spiral
opening angle of $\sim$0.2\,au degree$^{-1}$ should be considered to be
an upper limit, because the outer disk structure more closely resembles a circular ring when we use the $r^{1.35}$ height profile for
the deprojection, as we discussed at the end of Sec. 4.2. Hydrodynamical modeling of disks perturbed by embedded companions often show single-armed spirals both inside and outside the
companion’s orbital radius ($r_{\rm c}$). The spiral opening angles in the
model disks outside $r_{\rm c}$ either remain constant for a large range $r$
(\citealt{Ogilvie2002, Ragusa2017}) or opening angles
decreasing with $r$ (\citealt{Dong2015a}).\\
For $r < r_{\rm c}$ , the aforementioned models all predict spiral
opening angles that are larger than outside $r_{\rm c}$ . The under-sampling in the south-south-western (inner white-dashed) region
does not allow us to determine confidently how the pattern winds
inwards from the innermost point of the green dashed line. However, we do detect a sudden increase in spiral opening angle at
PA$\sim270^\circ$, $r\sim$200\,au (transition from green-dashed into red-dashed line). This change in spiral opening angle does allow the
option where feature B1 in the south-east is connected to the spiral pattern of the outer disk through a more complex spiral pattern with radially varying opening angle (red-white-red dashed
line). Feature A3 is much more diffuse after the deprojection, but
it does not seem possible to be explained with the simple spiral
of Eq.6.

\subsection{Planet detection limits}

Fig. 6 shows a map of the detection limits in the H2 filter, based
on an angular differential imaging (ADI, \citealt{Marois2006})
reduction of the SHINE data. The mass limits are based on
the DUSTY models (\citealt{Chabrier2000}). At large separations
$r \gtrsim 1 \arcsec{}$ , we reach an upper limit of $\sim 5 M_{\rm Jup}$ . However, the ADI
routine is strongly affected by the large azimuthal asymmetries in the disk, which dominates the noise at smaller separations.
Furthermore, if a planet orbits the star at these smaller separations near the disk mid-plane, its thermal emission will be heavily extincted due to scattering and absorption of the surrounding
circumstellar disk. Therefore, we cannot conclude from these observations that there is no companion of mass higher than
shown in Fig. 6 orbiting within 1\arcsec{} from the star. However, because neither our SPHERE polarimetric images, nor the ALMA
continuum images of \cite{vanderPlas2017} reveal any signal
from the dust disk at $r \gtrsim 1 \arcsec{}$ , we do consider the $\sim 5 M_{\rm Jup}$ limit
at larger separations to be reliable.

\begin{figure}
\center
\includegraphics[width=0.48\textwidth]{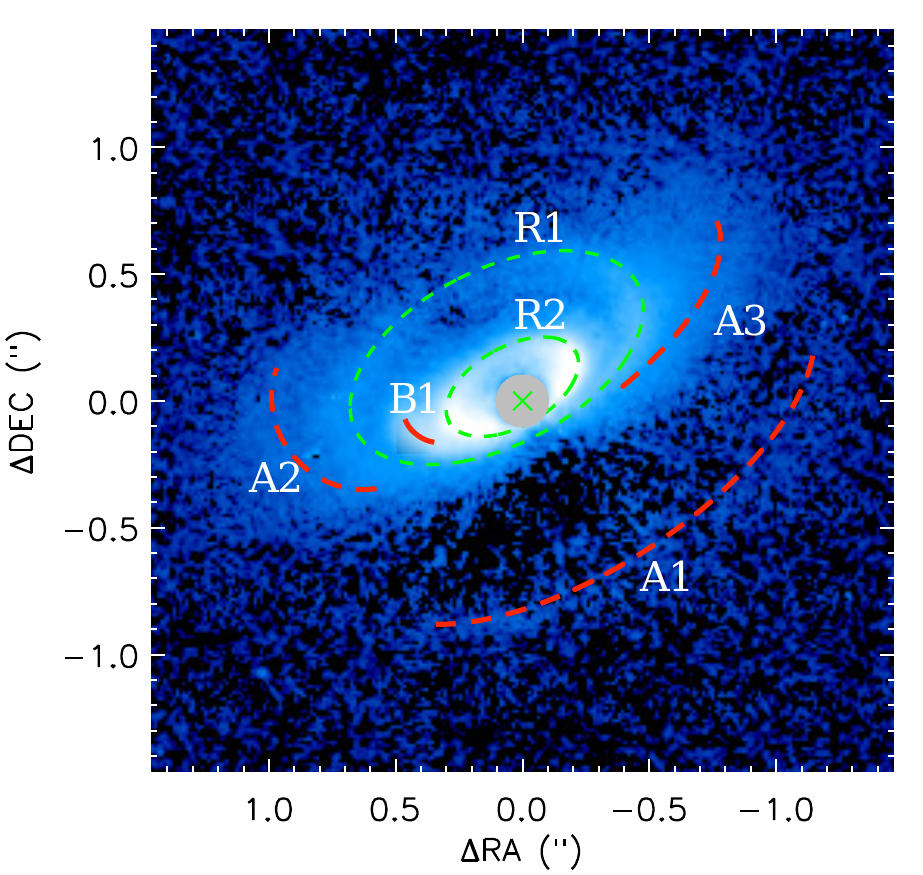} 
\caption{Same image as Fig. 1a, but displayed in logarithmic scale and
overlaid with arcs and with the most prominent disk features highlighted. We distinguish two possible rings (R1 and R2, green dashed ellipses), three arcs (A1, A2 and A3, red dashed arc) and a small arc or
blob in between the two rings (B1, red solid arc). While the red arcs
are only highlighting the features mentioned above, the green dashed
ellipses are showing the best fits for R1 and R2 with the fit parameters
listed in Tab. 1. The grey circle at the location of the star shows the size
of the coronagraph inner working angle.
} 
\label{fig:hd34282-sphere-overlay}

\end{figure}

\begin{figure*}
\center
\includegraphics[width=0.98\textwidth]{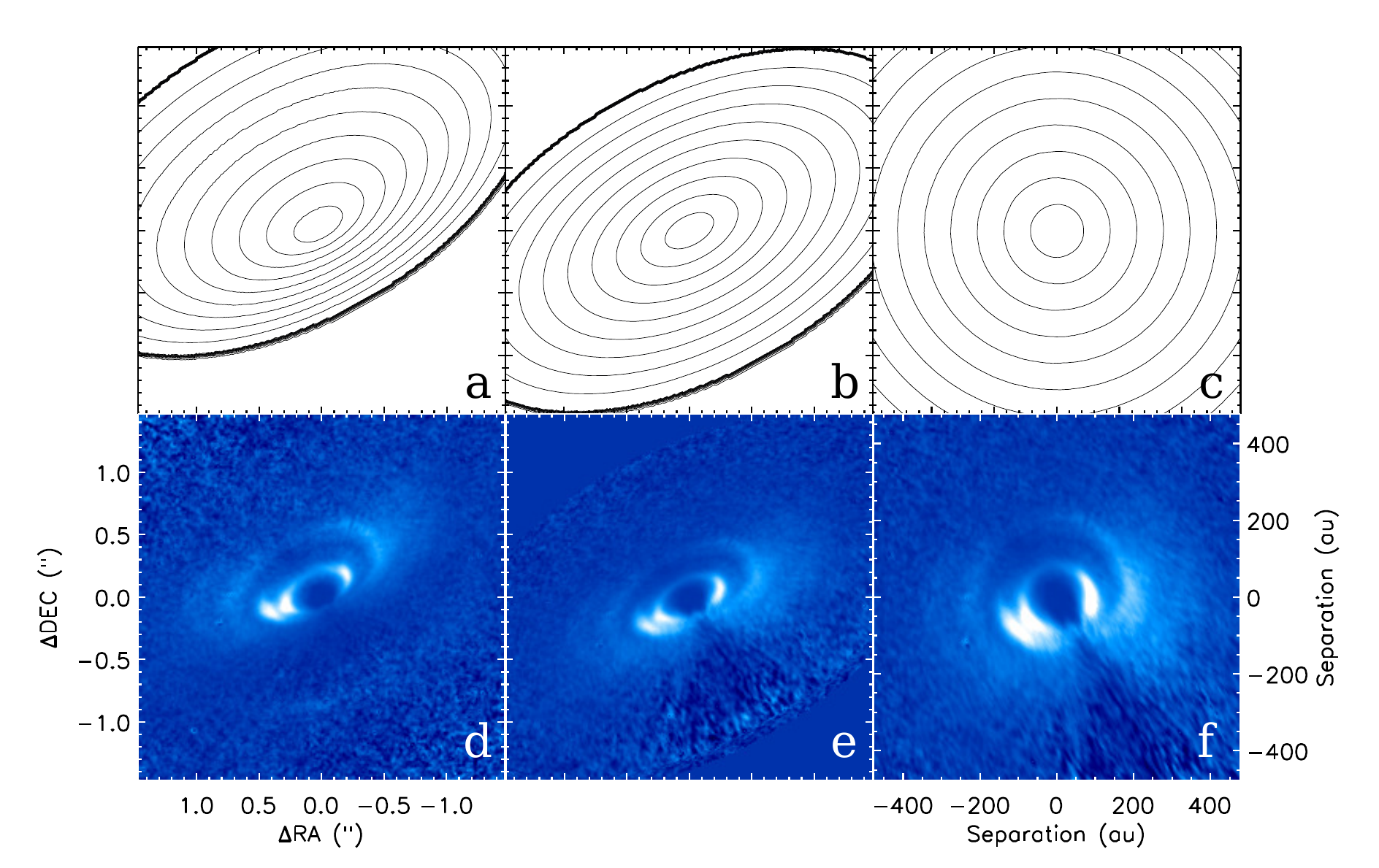} 
\caption{[a] Map of the height profile using the function H scat (r) = 0.29 r. [b] Height map shifted with $H_{\rm scat} \times$ sin i in the direction of the minor axis
to create concetric ellipses. [c] The centered height map of panel b is stretched along the minor axis with 1 / cos i. [d] Original inclination corrected
r 2 scaled Qphi image (as Fig. 1) c. [e] Qphi image shifted in the same way as the height map in panel b. This step creates a ‘flattened’ but inclined
image. [f] Stretched similar to the deprojection of panel c, to show the disk similar to a face-on orientation.
} 
\label{fig:hd34282-sphere-deprojected}

\end{figure*}

\begin{figure*}
\center
\includegraphics[width=0.98\textwidth]{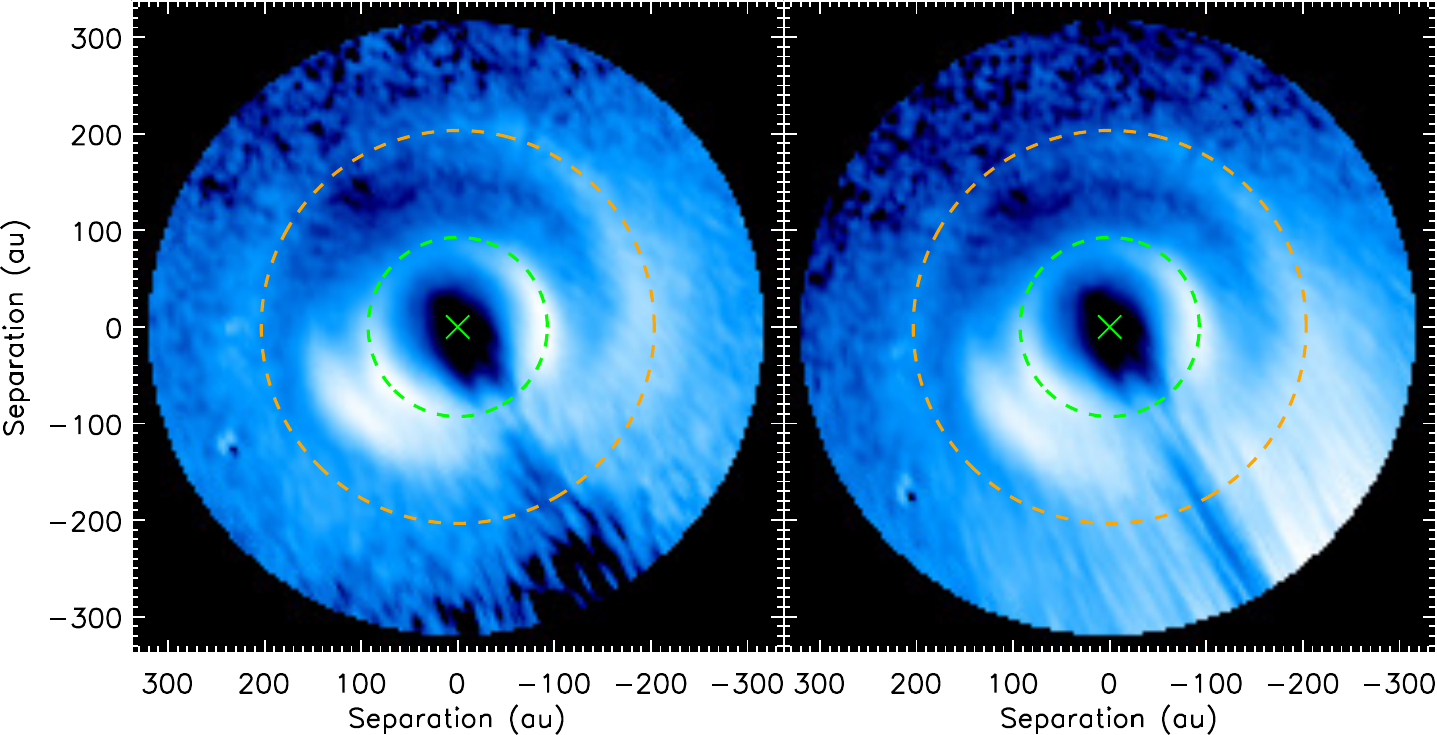} 
\caption{[a] Height-corrected deprojection using the non-flaring height profile (Eq. 5). This deprojection is the same as for Fig. 3f, but is displayed
here in log scale (not r 2 scaled). [b] As panel a, but deprojected with the flaring height profile (Eq. 4).
} 
\label{fig:hd34282-sphere-deprojected-overlay}

\end{figure*}

\begin{figure}
\center
\includegraphics[width=0.48\textwidth]{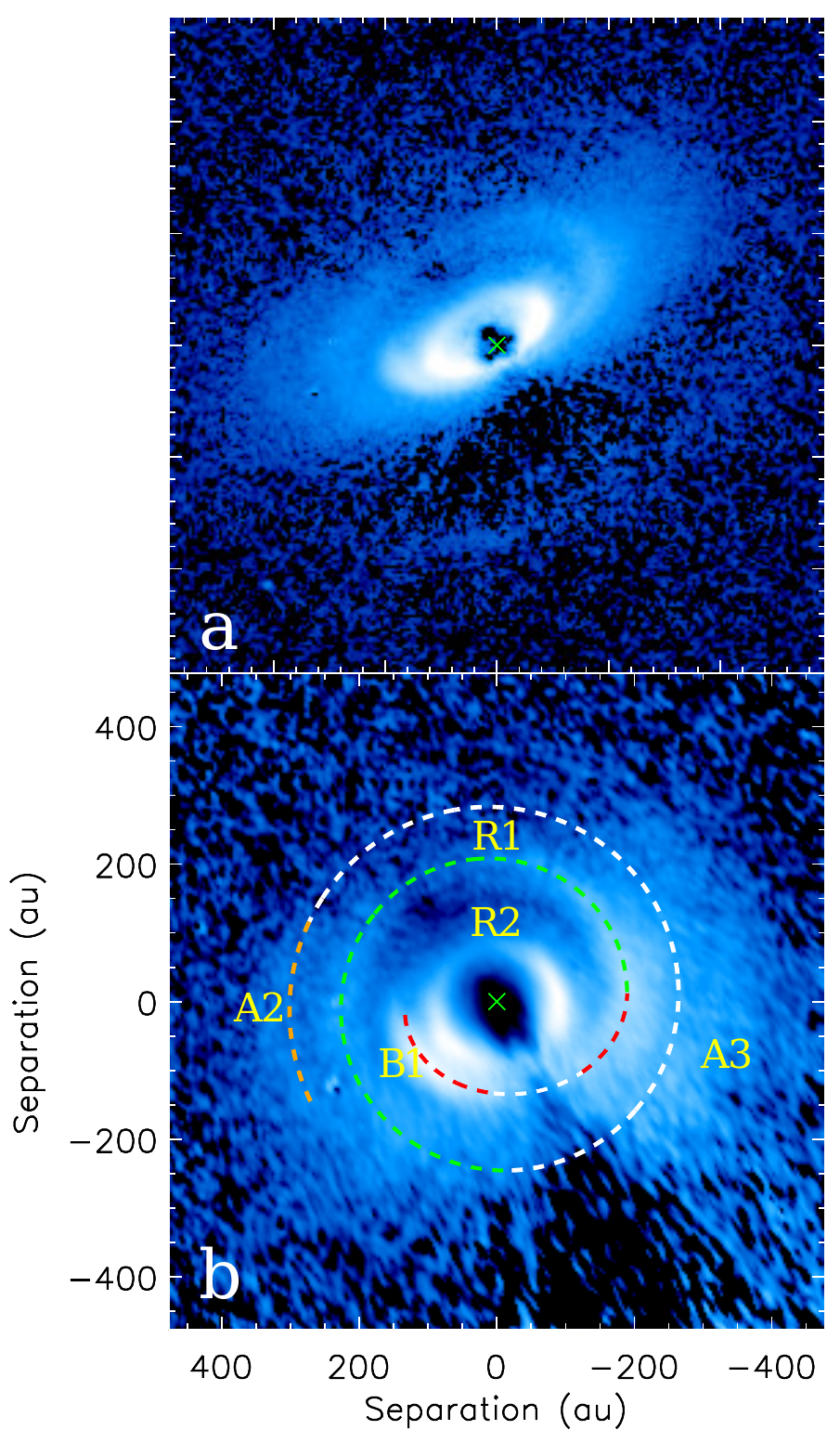} 
\caption{[a] [a] For comparison, we show the log scaled Qphi image again
without highlighted features. The tickmarks are the same as for Fig. 2.
[b] Spiral arm overlayed on the deprojected image (log scale) using the
non-flaring height profile, with the same feature annotations as used
in Fig. 2. The outer green-white-orange spiral (containing features R1
and A2) is a simple Archimedian spiral as described by Eq. 6, while the
inner red-white-red dashed line is a mere illustration of how B1 could be
connected to the base of the green spiral. The green and orange dashed
lines trace the R1 and A2 structures, respectively. The red dashed lines,
which trace the inner structures to the south east and south west of the
star clearly have different spiral properties than the outer spiral, most
notably at the transition from red to green. Although no clear radial
peak is detected at the positions of the white dashed lines in both the
outer and inner structures, they follow the same behaviour as the outer
and the inner arms, respectively. Feature A3 is difficult to recognize
because it is stretched into a diffuse region outside the white spiral, and
is not described by any dashed line.
} 
\label{fig:hd34282-sphere-comparison}

\end{figure}

\begin{figure}
\center
\includegraphics[width=0.48\textwidth]{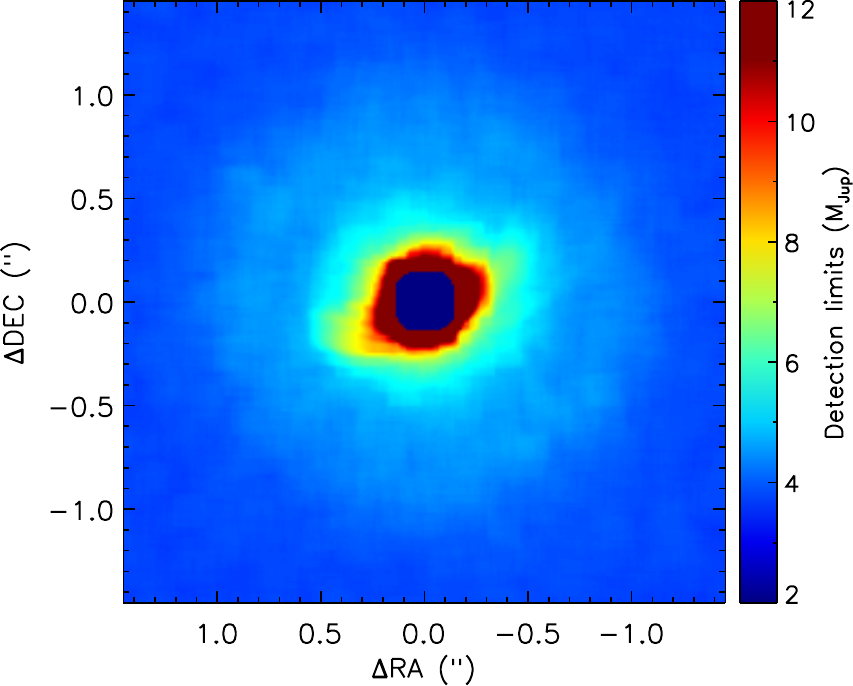} 
\caption{Detection map of angular differential imaging result in the
H2 filter, expressed in Jupiter masses based on the DUSTY models
\citep{Chabrier2000}. Notice that the region containing disk signal contains $\sim$ 3-4 times higher upper limits than the background, due to noise added to the post-processing routine by the disk signal. } 
\label{fig:hd34282-sphere-detection-map}

\end{figure}

%--------------------------------------------------------------------
%Discussion
%--------------------------------------------------------------------

\section{Discussion}

\subsection{Single-armed spiral pattern in the outer disk}

For disks where it is possible to determine a first-order estimate of the height profile, the height-corrected de-projection we
present in Sec. 4.2 offers a useful tool to both further constrain
the shape of the height profile and to analyze whether disk features are circular or deviate from this shape. Although it cannot
fully be ruled out that feature R1 is a circular ring, Figs. 4 and 5 b
show that a single-armed spiral nature of this feature is more plausible.\\
Spiral density waves in the gas are a common prediction of hydrodynamic models of planet-disk interaction. Typically multiple spiral arms are excited inside of the planet location, while outside of the planet location single armed spirals are possible (\citealt{Zhu2015, Dong2015a, Fung2015, 2017ApJ...835...38D, Miranda2019}).
If we assume that the spiral structure seen in HD\,34282 is caused by a single planet then this leaves us with two main scenarios.
Either the planet is located close to the northern tip of feature B1, i.e., the structure we trace is entirely located outside of the planetary orbit, or the planet is located further out but the resulting spiral arms are too tightly wound to be resolved in the SPHERE observations.\\
The first scenario would make the feature that we trace in the outer disk a "true" single armed spiral. The shape and contrast of spiral features will depend on the precise parameters of the disk and the perturber. \cite{Dong2015a} show that inner and outer spiral features will vary significantly based on the mass of the pertuber, its location in the disk and the thermal properties of the disk (i.e., they investigate the extreme isothermal and adiabatic cases). For higher planet masses (they use 6\,M$_{\rm Jup}$) the outer arm in the disk is typically not well visible. This is similar to later results in \cite{Dong2016}, where it is shown that a low mass stellar companion will not excite detectable outer arms. 
Similarly, companions at smaller separations to the primary star will drive more prominent outer spiral arms (\citealt{Dong2015a}).
However, across different models the inner spiral arms are typically significantly brighter than the outer spiral arm. This is not well consistent with our data since we presumably would predict the outer spiral but not the inner ones. Nonetheless it is possible that the perturbing planet is located close to the inner disk in the case of HD\,34282, consistent with a prominent outer spiral arm. In that case it may be that we lack the resolution and sensitivity to pick up the inner arms close to the coronagraphic mask.\\
In the second scenario the planet may be located anywhere along the spiral feature in principle. One plausible location would be the discontinuity between the innermost part of the feature we trace from the South-East to the South-West and the outer part of the structure, as discussed in section~\ref{sec: spiral-description} (the connection point of the red-white dashed line and the green-dashed line in figure 5b). The discontinuity might indicate the "shoulder" typically visible at the planet location in hydrodynamic simulations. This scenario is similar to the one displayed in \cite{Dong2016} for a 3\,M$_{\rm Jup}$ planet under a 50$^\circ$ inclination and a position angle of 150$^\circ$ (see their figure 8). We note however that we do not see a clear connection of feature B1 to the North-Eastern part of the disk, as is visible in the Dong et al. models. 
Such a placement of the planet would in any case make the bright feature B1 and its continuation, marked with a red-dashed line in figure~5b, inner spiral structures. The remaining (fainter) structures that are traced in 5b would then be part of an outer spiral arm. This brightness distribution is consistent with the discussed models. \\
In either scenario we have to take into account that lower planet masses will typically produce more tightly wound spirals (see, e.g., \citealt{Dong2015a}). Thus the lower the planet mass, the harder a singled armed spiral is distinguishable from multi-armed spirals. This is consistent with the SPHERE detection limits shown in figure~6, which give a tentative upper limit of 6-7\,M$_{\rm Jup}$ at the suspected planet locations (with the caveats mentioned in section~4.4).

\subsection{Comparison with ALMA continuum image}

The ALMA data of \cite{vanderPlas2017} (Fig. 7) do not display a spiral pattern, rather a single broad annulus with an inner
cavity consistent with our observations. The otherwise smooth
annulus in the continuum image does display an increase in
surface brightness at a similar radius and (slightly larger) position angle as our B1 feature. This approximate overlap between the NIR and sub-mm asymmetric features supports the scenarios suggested by van der Plas et al. that we have either detected
a local increase in temperature, or that we are seeing a horseshoe induced by either a vortex or an over-density in the gas
disk (\citealt{Ragusa2017}), similar to the horse-shoe features detected in the disks of MWC 147 (\citealt{vanderMarel2013}) and
HD 142527 (\citealt{Casassus2013}).\\
Recent modeling work of \cite{Dong2017} shows the simultated H-band images of a disk containing a super-earth that
display a faint trace of a single-armed spiral pattern (although the
detectability is questioned by the authors), while the simulated
sub-mm images are not sensitive to the spiral pattern and only
show a gap. Long baseline ALMA observations of HD 34282
are required to definitively determine whether there is a spiral
pattern and/or a gap (outside the inner rim of the disk, possibly
coincident with the dark region we detect at $\sim$150 au) in the sub-mm dust. We leave the detailed analysis of the spiral pattern and
the possibility to derive properties of a potential protoplanet for
future work where we examine ALMA continuum images and
these SPHERE data with radiative transfer modeling.

\begin{figure}
\center
\includegraphics[width=0.48\textwidth]{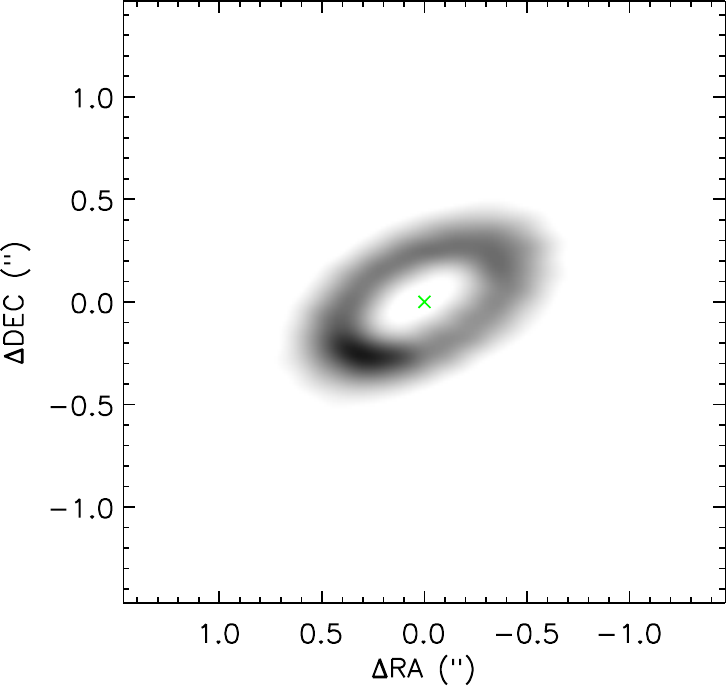} 
\caption{ALMA band 7 continuum image \citep{vanderPlas2017}. The disk is detected in as an annulus, with an azimuthal asymmetry at PA=135  . Contrary to the SPHERE detection of the scattering surface, the sub-mm image probes the mid-plane of the disk. Therefore, the annu-lus detected with ALMA does not show a height-induced offset, but is centered around the star.
} 
\label{fig:hd34282-alma}

\end{figure}

\section{Conclusions}

The first detection of the protoplanetary disk around HD 34282
in scattered-light reveals a disk surface rich in structure. In accordance with the model predictions of \cite{Khalafinejad2016}
the disk contains a cavity surrounded by the disk’s puffed-up inner rim at 88\,au. Because of the inclination ($i = 56^\circ$) of the system, the disk displays indications of its large vertical extent: the
top side of the disk appears offset from the star-center while the
bottom side is clearly visible with an opposing offset.\\
We create two surface height-profiles: one using a constant
$H_{\rm scat} /r = 0.29$, the other using a power law $\propto r^{1.35}$ . Both profiles
are used to perform a height-corrected deprojection or ‘rotation’
of the disk to simulate face-on images of the disk. This deprojection shows evidence that the disk region outside the puffed-
up inner rim (R2) contains a tightly wound single-armed spiral,
winding outwards in counter-clockwise direction. Although the outermost region can be approximated by the Archimedian spiral description of Eq. 6, the region inward of r$\sim$180 au cannot
be described in such simple terms.\\
In future work, we will continue investigating HD 34282 system by performing radiative transfer modeling including both
the ALMA data of \cite{vanderPlas2017} and the
presented SPHERE data. Subsequently, we will observe the disk
with ALMA with a longer baseline to obtain sub-mm images of
a resolution comparable to that of our SPHERE observations, to
search for substructure in the annulus detected by van der Plas
et al.

\begin{acknowledgements}

We thank Ruobing Dong for our discussions on the appearance
of inclined spiral systems. JdB acknowledges the funding by the
European Research Council under ERC Starting Grant agreement 678194 (FALCONER).
SPHERE is an instrument designed and built by a consortium
consisting of IPAG (Grenoble, France), MPIA (Heidelberg, Germany),
LAM (Marseille, France), LESIA (Paris, France), Laboratoire Lagrange
(Nice, France), INAF - Osservatorio di Padova (Italy), Observatoire de
Genève (Switzerland), ETH Zurich (Switzerland), NOVA (Netherlands), ONERA
(France), and ASTRON (The Netherlands) in collaboration with ESO.
SPHERE was funded by ESO, with additional contributions from CNRS
(France), MPIA (Germany), INAF (Italy), FINES (Switzerland), and NOVA
(The Netherlands). SPHERE also received funding from the European Commission
Sixth and Seventh Framework Programmes as part of the Optical Infrared
Coordination Network for Astronomy (OPTICON) under grant number RII3-Ct2004-001566
for FP6 (2004-2008), grant number 226604 for FP7 (2009-2012),
and grant number 312430 for FP7 (2013-2016). 

\end{acknowledgements}

\bibliographystyle{aa}
\bibliography{MyBibFMe.bib}
%
%\newpage
\begin{appendix}

\section{Phase-function variations}

The resulting height-corrected deprojections, shown in Fig. 4a \&
b simulate the rotation of the disk towards a face-on ($i = 0^\circ$) orientation. The main difference with a truely face-on observation
of the disk is that the large azimuthal variation of the scattering
angles (always $\sim$90$^\circ$ for a truely face-on disk) are not corrected
for. While a homogeneous ring centered around the star would
not vary in surface brightness for a truely face-on disk, it does
still vary in azimuthal direction after our height-corrected deprojection, due to variations in the scattering phase-function. To
illustrate this effect, we have created a simple disk ring seen at
$i = 56^\circ$ (Fig. A.1 a) with the radiative transfer code MCFOST
(\citealt{Pinte2006}). Subsequently, after setting the central 8 pixels to zero to mimic the presence of a coronagraph mask, we
have rotated the image using the same height-corrected deprojection as used in Fig. 3. Finally, we have smoothed the image
with a Gaussian of Full Width at Half Maximum of 3 pixels.
This deprojected model image, shown in Fig. A.1 b displays similar variations in surface brightness along the ring: bright at small
(forward) scattering angles and faint for large (backward) scattering angles.\\
Furthermore, it is important to note the effect of the deprojection on the bottom side of the ring. The shift to reverse the
offset u of the scattering surface (as shown in the second column of Fig. 3) is only performed in one direction: towards the
south-west. However, the bottom side of the disk has an offset in
the direction opposite to the top surface and is therefore shifted
in the wrong direction (even further away from the star-center).
Although the bottom of ring R2 is not detected for HD\,34282,
feature A1 most likely represents the bottom of the outer regions (R1 or even further out) of the disk and is therefore shifted too
far away from the center in our deprojection as well (outside the
frame of Fig. 3f). However, the aim of the height-corrected deprojection is to simulate a face-on orientation, at which it would
be completely impossible to see the bottom side (for an optically thick disk). Since signal from the bottom regions (A1) of
the disk is shifted away from the top surface we have avoided
contamination of the top surface.

\begin{figure}[!ht]
\center
\includegraphics[width=0.48\textwidth]{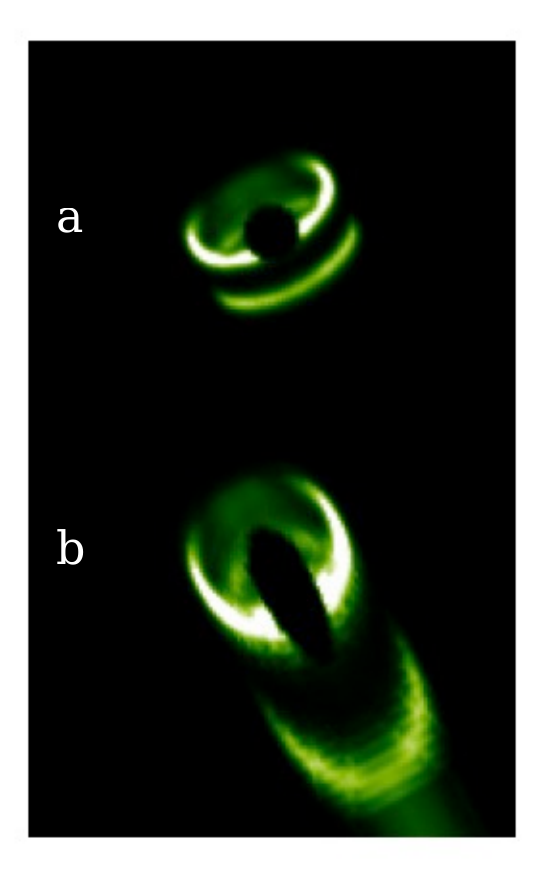} 
\caption{[a] Raditative transfer (MCFOST) disk-model comprised of
a single ring with similar i and H scat /r as the disk of HD 34282. A software mask is added to the star-center of the image to simulate the presence of a coronagraph mask. Finally the image is smoothed with a Gaussian. [b]
The model image of panel [a] after a height-corrected deprojection. Note
that this method shifts the bottom side of the disk (arc in bottom-right
corner of the image) in the wrong direction.
} 
\label{fig:appendix}

\end{figure}

\end{appendix}

\end{document}